\documentclass[trackchanges, twocolumn]{aastex7}

\newcommand\R[1]{\textbf{#1}}

\begin{document}

\title{Flares on TRAPPIST-1 reveal the spectrum of magnetic features on its surface}

\author[0009-0009-3020-3435]{Valeriy Vasilyev}
\affiliation{Max Planck Institute for Solar System Research, Justus-von-Liebig-Weg 3, 37077 G\"ottingen, Germany
}
\email[show]{vasilyev@mps.mpg.de }  
\author[0000-0002-6087-3271]{Nadiia Kostogryz}
\affiliation{Max Planck Institute for Solar System Research, Justus-von-Liebig-Weg 3, 37077 G\"ottingen, Germany
}
\email[]{}  
\author[0000-0002-8842-5403]{Alexander I. Shapiro}
\affiliation{ Institute of Physics, University of Graz, 8010 Graz, Austria
}
\affiliation{Max Planck Institute for Solar System Research, Justus-von-Liebig-Weg 3, 37077 G\"ottingen, Germany
}
\email[]{}

\author[0000-0003-2073-002X]{Astrid M. Veronig}
\affiliation{ Institute of Physics, University of Graz, 8010 Graz, Austria
}
\affiliation{University of Graz, Kanzelh\"ohe Observatory for solar and Environmental Research, Kanzelh\"ohe 19, 9521 Treffen, Austria
}
\email[]{}

\author[0000-0002-3627-1676]{Benjamin V. Rackham}
\affiliation{Department of Earth, Atmospheric and Planetary Sciences, Massachusetts Institute of Technology, Cambridge, MA 02139, USA}
\affiliation{Kavli Institute for Astrophysics and Space Research, Massachusetts Institute of Technology, Cambridge, MA 02139, USA}
\email[]{}

\author[0009-0004-2182-2596]{Christoph Schirninger}
\affiliation{ Institute of Physics, University of Graz, 8010 Graz, Austria
}
\email[]{christoph.schirninger@uni-graz.at}

\author[0000-0003-2415-2191]{Julien de Wit}
\affiliation{Department of Earth, Atmospheric and Planetary Sciences, Massachusetts Institute of Technology, Cambridge, MA 02139, USA}
\email[]{}

\author[0000-0002-0583-0949]{Ward Howard}
\affiliation{Department of Astrophysical and Planetary Sciences, University of Colorado, 2000 Colorado Avenue, Boulder, CO 80309, USA
}
\affiliation{NASA Hubble Fellowship Program Sagan Fellow
}
\email[]{}

\author[0000-0003-3305-6281]{Jeff Valenti}
\affiliation{Space Telescope Science Institute, 3700 San Martin Drive, Baltimore, MD 21218,
USA}
\email[]{}

\author[0000-0002-9464-8101]{Adina D. Feinstein}
\affiliation{Department of Physics and Astronomy, Michigan
State University, East Lansing, MI, USA
}
\email[]{}

\author[0000-0003-4676-0622]{Olivia Lim}
\affiliation{Institut Trottier de recherche sur les exoplan\`etes, Université de Montréal, 1375 Ave Thérèse-Lavoie-Roux, Montréal, QC, H2V 0B3, Canada}

\email[]{}

\author[0000-0002-6892-6948]{Sara Seager}
\affiliation{Department of Physics,
Massachusetts Institute of Technology, Cambridge, MA 02139, USA}
\affiliation{Department of Earth, Atmospheric and Planetary Sciences, Massachusetts Institute of Technology, Cambridge, MA 02139, USA}
\affiliation{Kavli Institute for Astrophysics and Space Research, Massachusetts Institute of Technology, Cambridge, MA 02139, USA}
\affiliation{Department of Aeronautics and Astronautics,
MIT, 77 Massachusetts Avenue, Cambridge, MA 02139, USA}
\email[]{}

\author[0000-0001-7696-8665]{Laurent Gizon}
\affiliation{Max Planck Institute for Solar System Research, Justus-von-Liebig-Weg 3, 37077 G\"ottingen, Germany
}
\affiliation{Institut für Astrophysik, Georg-August-Universit\"at G\"ottingen,  37077  G\"ottingen, Germany
}
\email[]{}

\author[0000-0002-3418-8449]{Sami K. Solanki}
\affiliation{Max Planck Institute for Solar System Research, Justus-von-Liebig-Weg 3, 37077 G\"ottingen, Germany
}
\email[]{}

\begin{abstract}
TRAPPIST-1 is an M8 dwarf hosting seven known exoplanets and is currently one of the most frequently observed targets of the James Webb Space Telescope (JWST). However, it is notoriously active, and its surface is believed to be covered by magnetic features that contaminate the planetary transmission spectra. The radiative spectra  of these magnetic features are needed to clean transmission spectra, but they currently remain unknown. 
Here, we develop a new approach for measuring these spectra using time-resolved JWST/NIRISS observations. 
We detect a persistent post-flare enhancement in the spectral flux of TRAPPIST-1. Our analysis rules out lingering flare decay as the cause of the flux enhancement and, thus, points to structural changes on the stellar surface induced by flares. We suggest that the flaring event triggers the disappearance of (part of) a dark magnetic feature, producing a net brightening. This suggestion is motivated by solar data: flare-induced disappearance of magnetic features on the solar surface has been directly detected in high spatial resolution images, and our analysis shows that this process produces changes in solar brightness very similar to those we observe on TRAPPIST-1. The proposed explanation for the flux enhancement enables, to our knowledge, the first measurement of the spectrum of a magnetic feature on an M8 dwarf. Our analysis indicates that the disappearing magnetic feature is cooler than the TRAPPIST-1 photosphere, but by at most a few hundred kelvins.
\end{abstract}

\keywords{\uat{Starspots}{1572} --- \uat{Stellar activity}{1580}  --- \uat{Stellar flares}{1603}}

\section{Introduction} 
\label{sec:intro}
Since the discovery of seven terrestrial planets orbiting TRAPPIST-1 \citep{Gillon2017, Luger2017}, this nearby M-dwarf has become one of the most intensively studied main-sequence stars after the Sun. Notably, three of its planets reside within the habitable zone, making the system a prime target for multiple JWST programs aimed at detecting and characterizing planetary atmospheres. However, a major challenge of this effort has been the notorious magnetic activity of TRAPPIST-1 \citep{Ducrot2018, Wakeford2019, Garcia2022}. The transmission spectra of its planets vary significantly between transits \citep{Lim2023}, complicating the identification of atmospheric signals. These variations are thought to arise from unocculted magnetic features on the surface of TRAPPIST-1 \citep{Rathcke2025}. Attempts to model and correct for these effects are hindered by the lack of constraints on the properties of these features \citep{Davoudi2024,TRAPPISTInitiative2024,Radica2025}. To date, no unambiguous spot-crossing events have been observed in the system, and no simulations of TRAPPIST-1’s magnetic features have been published.

Here, we propose an intriguing avenue for indirectly constraining the spectra of magnetic features on TRAPPIST-1 by leveraging another magnetic phenomenon observed in all JWST programs targeting the TRAPPIST-1 system: flares, which manifest as sudden brightness enhancements of the star \citep[e.g.,][]{Howard2023}. We report a stable post-flare increase of TRAPPIST-1 flux that persists after the disappearance of emission lines associated with the flare. 
Such an increase can be reliably identified in all cases where data are available for at least 1.5 hours after the flare, allowing sufficient time for the flare-associated emission to subside. 
We interpret the  observed post-flare flux enhancement as evidence for a structural change on the stellar surface that happened during the flaring event, such as the disappearance of a dark magnetic feature triggered by the flare. 

This interpretation is inspired by solar observations that show a strong link between magnetic features and flares. Magnetic fields emerge from the solar interior, suppressing near-surface convection and giving rise to surface (photospheric) magnetic features \citep[see reviews by][]{Solanki2003, Solanki2006, Stein2012}. The reconnections of magnetic field in the solar corona lead to flares \citep[see][for reviews]{Priest2002,Shibata2011} and a reconfiguration of the surface magnetic field. This reconfiguration is sometimes accompanied by the disappearance of observable portions of photospheric magnetic features \citep[e.g.,][]{Wang2002, Deng2005}.

Although the disappearance of magnetic features following flares can be identified in solar images, such events are obscured in the Sun-as-a-star observations (i.e., in the disk-integrated solar spectra) by variations arising from the simultaneous evolution of other magnetic structures across the visible solar disk. In contrast, TRAPPIST-1 presents a markedly different case: its surface area is roughly 70 times smaller and its temperature more than twice as low, resulting in a luminosity just 0.05\% that of the Sun.  Yet, the flare energy–frequency distribution of TRAPPIST-1 closely resembles that of the Sun \citep{Vida2017, Seli2021, Feinstein2022}, suggesting that flares can produce proportionally much larger surface changes than those typically seen on the Sun. This offers an  opportunity to study the disappearance of magnetic features on TRAPPIST-1 — and even to measure their spectra.

\section{Solar Observations of Penumbral Disappearance}\label{section_sun}
\begin{figure*}[!htb]
    \centering
     \includegraphics[width=1\textwidth]{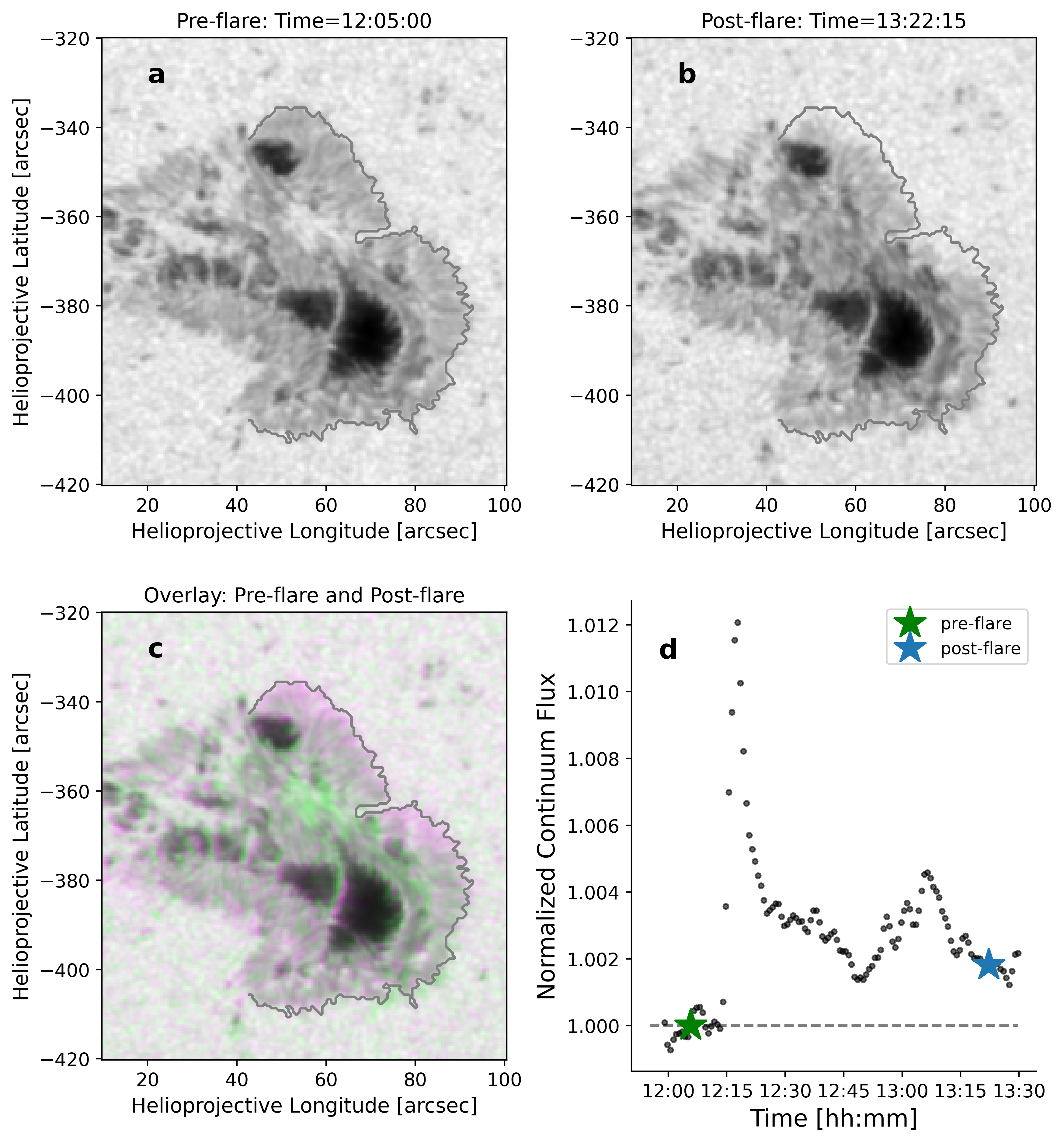}
\caption{\textbf{Disappearance of a part of the penumbra of a sunspot.} 
(a) Continuum image from SDO/HMI of an active region on the Sun taken before the X9.0 flare on 2024-10-03. The gray  contour outlines the Western penumbra to highlight the penumbral changes during the flare.  
(b) Continuum image of the same region after the flare, with the same gray contour overlaid to show the disappearance of part of the penumbra.
(c) Color overlay of the pre- and post-flare images (green: pre-flare; magenta: post-flare), with the same gray contour outlining the pre-flare penumbra. The magenta near the contour marks the area of the disappeared penumbral region.
(d)
Normalized continuum flux as a function of time on 2024-10-03, derived by integrating the flux over the full field of view shown in panel \textit{a}. The horizontal line indicates the mean pre-flare level calculated between 12:00--12:14 UT. The green and blue stars mark the times corresponding to the pre-flare and post-flare images shown in panels \textit{a} and \textit{b}. The flare is followed by a persistent enhancement of the continuum intensity. This figure demonstrates that a solar flare can cause (part of) an active region to vanish.
}
\label{fig:solar_observations}
\end{figure*}
We begin  by presenting a direct evidence that flare-induced disappearance of magnetic features can occur in nature. At present, such evidence is available only from the Sun, which serves as a unique natural laboratory, being the only star where we can spatially resolve and study  magnetic features associated with flares.

Changes and even disappearances of the sunspot penumbra during solar flares have been documented in the literature \citep{Wang2004, Deng2005, Xu2019}. In one case even the disappearance of a small sunspot was observed during a flare, and interpreted as an effect of the fast magnetic reconnection process and subsequent submergence of magnetic flux \citep{Wang2002}.

Here we use as an example observations of the strong X9.0 solar flare on 2024-10-03 that originated from Active Region 13842. The disappearance of sunspot penumbra associated with this flare has not yet been reported in the literature. Furthermore, we show for the first time that the flare-induced disappearance of the penumbral region leads to an enhancement of the post-flare spectral flux.

The 2024-10-03 flaring event was observed as a white-light flare in observations from the Helioseismic and Magnetic Imager \cite[HMI;][]{Scherrer2012} onboard the Solar Dynamics Observatory (SDO). HMI is an imaging spectropolarimeter that observes the full solar disk with a spatial resolution of 1$''$. Here we use the HMI continuum proxy derived from the Fe~{\sc i} line at 6173~{\AA}, which is measured at 6 wavelength points distributed  symmetrically around the nominal line center with a time cadence of 45 sec. The continuum data product has been used in several studies to detect and characterize solar white-light flares \cite[e.g.,][]{Kuhar2016,Namekata2017}. 

Figure~\ref{fig:solar_observations} and the accompanying movie\footnote{Available in the electronic supplemental materials.} show the evolution of the X9 flare in the SDO/HMI continuum images (see in particular the time frame 12:14 to 12:19 UT in the movie). The images have been corrected for solar differential rotation. In the movie, one can observe two elongated flare ribbons protruding into the sunspot and sweeping over a significant portion of the Western penumbra. One can also see a reconfiguration of the penumbra during the flare, where parts of the penumbra disappear or shrink. To better see the change,  we show the outline of the Western pre-flare penumbra in the two snapshots in Fig.~\ref{fig:solar_observations}a-b.

Figure~\ref{fig:solar_observations}d shows the emission integrated over the whole field-of-view shown in Fig.~\ref{fig:solar_observations}a, normalized by the mean pre-flare level (indicated by the horizontal line). 
One can clearly see the white-light flare during about 12:14 to 12:19~UT, but also notice an enhanced level of emission after the flare until the end of the plotted time series  (13:30~UT). We relate this enhancement to the sudden reconfiguration and disappearance of part of the sunspot penumbra that was swept by the western flare ribbon (see Fig.~\ref{fig:solar_observations}c). We emphasize that such an enhancement (along with the flux increase during the flare) can only be detected by integrating the continuum flux over a small surface area encompassing the active region. In contrast, the disk-integrated solar flux is influenced by the evolution of other active regions and solar granulation, which obscure the subtle signals from the flare and the even smaller post-flare flux enhancement. As a result solar flares are very rarely seen in the {\it disk-integrated} white-light \citep{Kretzschmar2010}.

While this disappearance is clearly visible in the continuum images, it is important to note that it does not necessarily imply the complete removal of the underlying magnetic field. 
Previous solar studies have shown that such morphological changes can instead result from a reconfiguration of the surface magnetic field, often involving rapid changes in orientation and a transition toward more horizontal fields during flares \citep[e.g.,][]{Wang2004, Deng2005, Sun2012}.
In this context, the observed disappearance should be understood as a reconfiguration of the magnetic structure’s orientation and its visible manifestation, rather than the disappearance of the magnetic field itself.

\section{JWST View of Magnetic Feature Disappearance on TRAPPIST-1}
Having illustrated the flare-induced disappearance of magnetic features using the exemplary case of the Sun, we now turn to the analysis of the TRAPPIST-1 observations. In Sect.~\ref{section_post_flare_flux} we present the post-flare flux enhancement in the JWST/NIRISS observations of TRAPPIST-1 that closely resembles solar observations.
Then, in Sects.~\ref{sec:spectral_features}–\ref{sec:correcltions}, we demonstrate that the post-flare flux enhancement is not a result of lingering flare decay and must therefore be caused by structural changes to the TRAPPIST-1 surface induced by flares. The most natural example of such structural changes, inspired by solar observations presented in Sect.~\ref{section_sun}, is the disappearance of (part of) a magnetic feature associated with the flare. By adapting such an interpretation of the post-flare flux enhancement we report in Sect.~\ref{sec:areas} the spectra of disappearing magnetic features and present approximate constraints on their sizes.

\subsection{Post-flare flux enhancements}\label{section_post_flare_flux}
\begin{figure*}[!htb]
    \centering
     \includegraphics[width=1.0\textwidth]{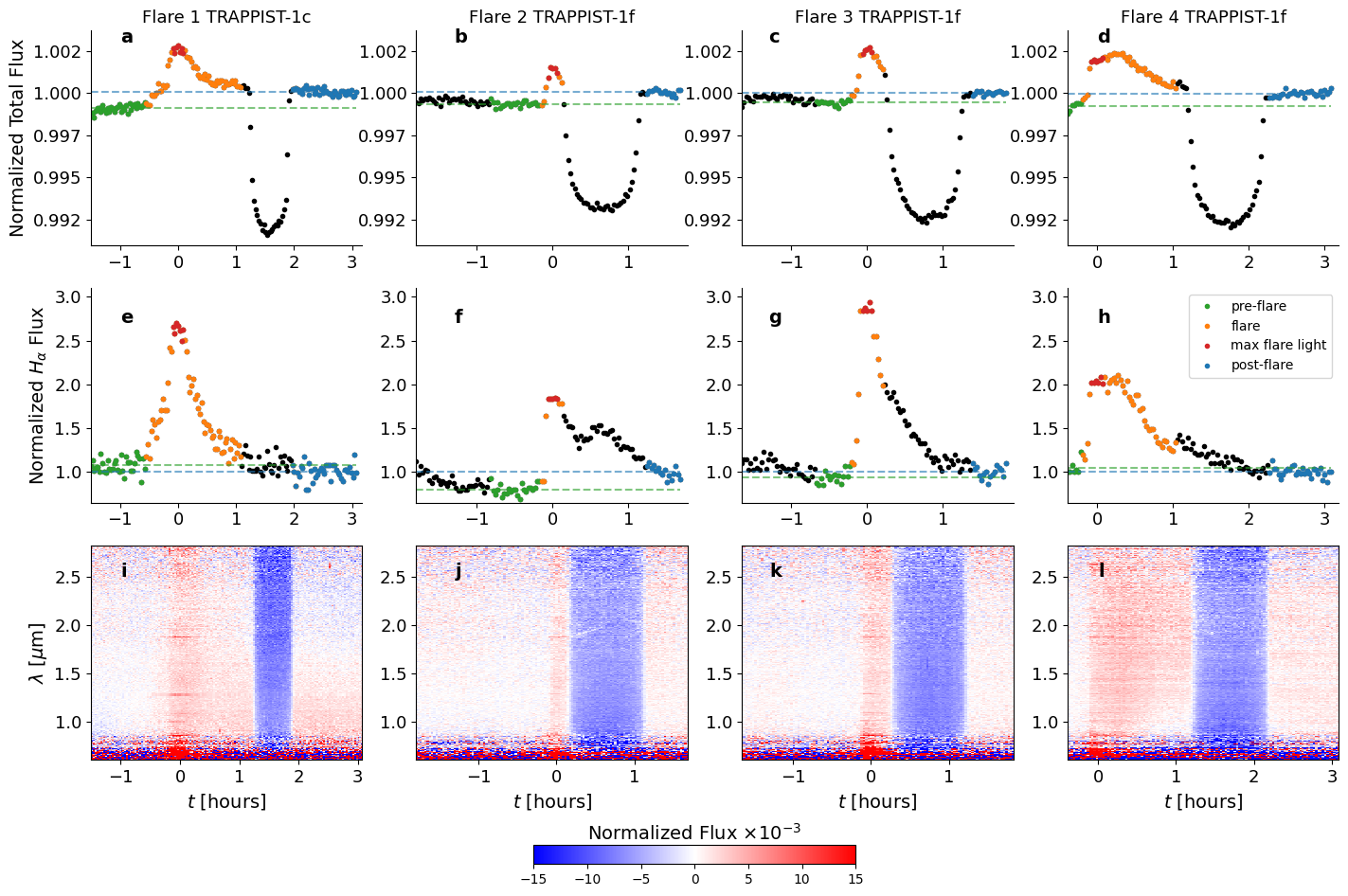}
\caption{\textbf{Stellar flares observed in JWST/NIRISS time-series spectroscopy of TRAPPIST-1.}
Each column shows one flare event during transits of the TRAPPIST-1c or TRAPPIST-1f planets.
\textit{Tow row:} Light curves with the total flux (integrated over the full spectral range and normalized to the mean out-of-transit flux). Colored points indicate pre-flare (green), flare (orange), maximum flare light (red), and post-flare (blue) phases. Green and blue dashed lines mark the average pre- and post-flare flux levels.
\textit{Middle row:} Light curves of the normalized H$\alpha$ flux.
\textit{Bottom row:} Time-resolved spectrograms (wavelength vs. time) normalized to the pre-flare spectrum. The shaded blue region shows the drop in flux caused by the planetary transit.
For comparison, the time axis  in all panels are aligned such that $t = 0$ corresponds to the beginning of the maximum flare light phase.  
In all four events, the total flux in the post-flare phase is systematically elevated relative to the pre-flare level.
}
\label{fig:datasets_and_offstes}
\end{figure*}
\begin{figure*}[!htb]
    \centering
     \includegraphics[width=1.0\textwidth]{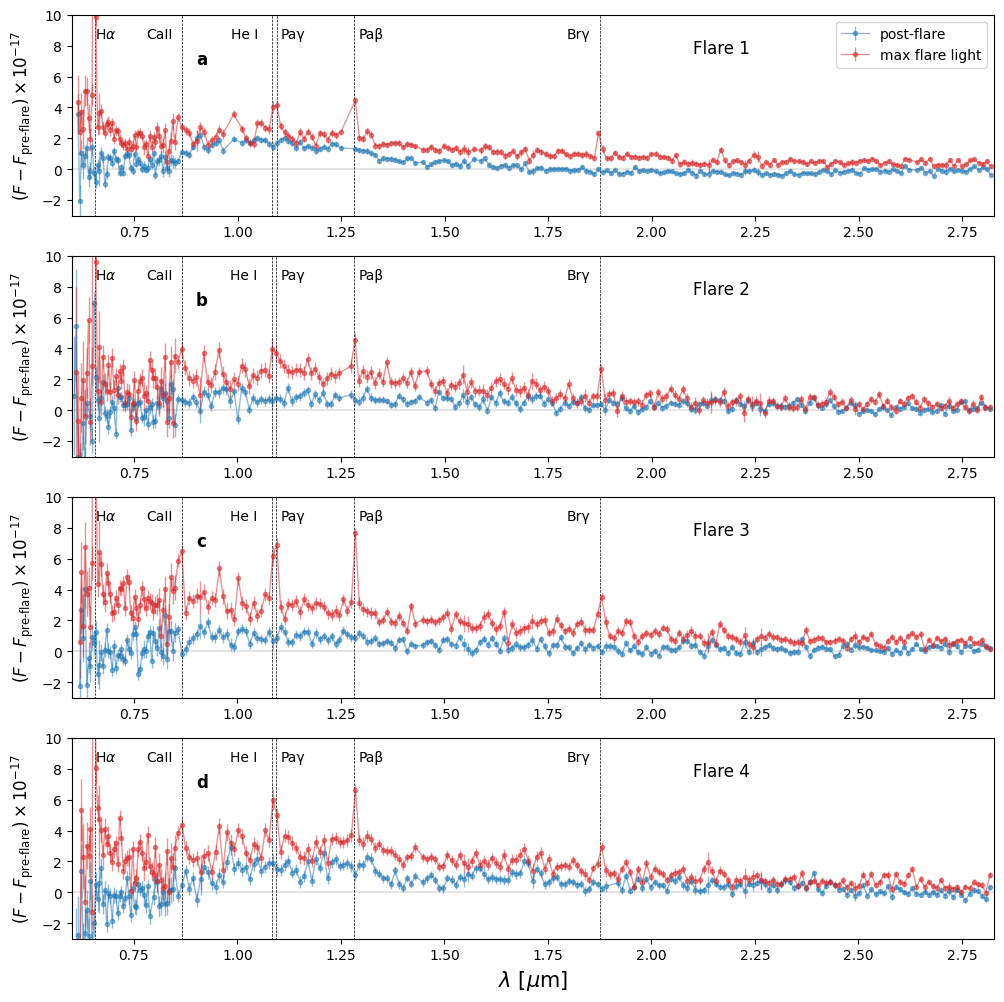}
\caption{\textbf{Spectral variations during and after the flare, relative to the pre-flare state.} 
\textit{Panels a-d:} Each panel shows one flare event (Flares 1–4).
The red curves represent the flare spectrum (maximum flare light minus pre-flare), and the blue curves show the post-flare spectrum relative to the pre-flare level, reflecting the spectral energy distributions of a disappearing active region. Vertical dashed lines indicate emission lines observed during the flare. All spectra are binned in wavelength by averaging over 20 points per bin for clarity. The spectral profile of the post-flare flux enhancement is substantially different from that observed during the flare maximum. In particular, it is not accompanied by the strong emission lines seen during the flare phase.
}
\label{fig:spectral_variations}
\end{figure*}
We analyzed four stellar flare events (hereafter Flare 1, 2, 3, and 4) captured by JWST/NIRISS time-series spectroscopy of TRAPPIST-1, obtained during planetary transit observations (see Appendix~\ref{app:data}). These datasets were selected based on the criterion that at least 1.5 hours after the flare maximum phase is available (as a result the flare occurred before the planetary transit in all selected datasets). Since all flares occurred prior to the planetary transits in the selected datasets, the transits overlap with the decay phase of the flare. As our analysis focuses on the post-flare phase, we exclude all in-transit data points and compute the post-flare flux levels using only data obtained after the transit. This approach minimizes contamination from both the transit signal and the flare decay.

In Figure~\ref{fig:datasets_and_offstes}a-b, we show light curves integrated over 0.6–2.8~$\mu$m range (hereafter referred as total flux light curves) for the four flare events.  For clarity, we divide each light curve into four distinct time windows: pre-flare (quiescent level before the event), maximum flare light (centered around the flare peak), the broader flare phase (including both rise and decay), and post-flare (after the event, once the light curve stabilizes).  Each flare exhibits a similar temporal evolution: a rapid rise in flux, a short-lived peak, and an exponential-like decay. The initial rise and final decay correspond to the impulsive and gradual phases described in stellar flare studies \citep[e.g.,][]{Davenport2014, Kowalski2024}. 
In all four events, the total flux in the post-flare phase is  systematically elevated relative to the pre-flare level. 

We also measured the flux in H$\alpha$ line (see panels e–h in Fig.~\ref{fig:datasets_and_offstes}) --  a well-established tracer of flaring magnetic activity on both the Sun and M dwarfs \citep[e.g.,][]{Benz2017, Kowalski2024}. We integrated the unbinned, non-normalized spectrum over a $\Delta \lambda \approx 0.43$~nm window centered on H$\alpha$, without continuum subtraction. As a check, we also separately measured the H$\alpha$ line and nearby continuum following \cite{Howard2023}, found that continuum subtraction does not affect the time evolution of the H$\alpha$ emission.

While the total flux enhancement persists after the flare in all four cases (see dashed lines in Fig.~\ref{fig:datasets_and_offstes}a–d), the H$\alpha$ flux returns to the pre-flare level in Flare 1, 3, and 4.  Notably, the Flare 1 dataset, which corresponds to the longest available post-flare observational period, shows a clear drop in H$\alpha$ flux in the post-flare phase below the pre-flare level (see Fig.~\ref{fig:datasets_and_offstes}e).  This is consistent with the vanished magnetic feature  being a source of H$\alpha$ emission.  In contrast, Flare 2 shows continued H$\alpha$ enhancement in the post-flare phase (see Fig.~\ref{fig:datasets_and_offstes}f), attributed to its shorter available post-flare observational period and the presence of a secondary flare event in the decay phase of the main flare.

We further illustrate the post-flare flux enhancement by presenting time-resolved spectrograms for all four flare events considered  in this study (Fig.~\ref{fig:datasets_and_offstes}i–l). Each spectrogram across the entire spectral range reveals a flux enhancement during the flare with a strong continuum increase at shorter wavelengths and with prominent emission lines (see Appendix~\ref{app:emission_lines} for details). These flare signatures decay across the entire spectral range on a timescale consistent with the light curves. However, after the flare subsides, we observe the flux enhancement visible across nearly all wavelengths. 

\subsection{Spectral features in the flare and the flux enhancement}\label{sec:spectral_features}
\begin{figure*}[ht!]
    \centering
     \includegraphics[width=1.0\textwidth]{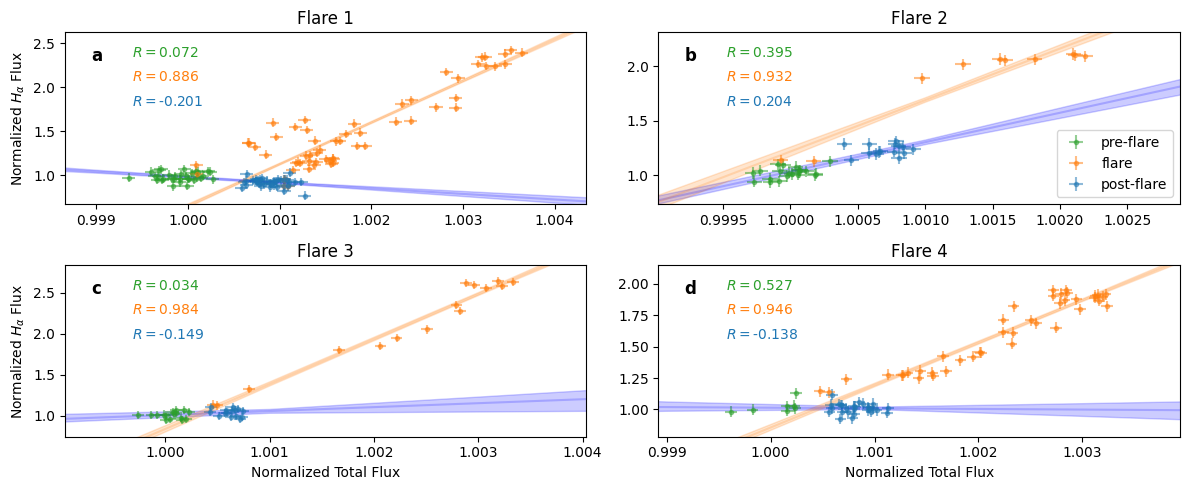}
\caption{\textbf{Correlation between total  and H$\alpha$ fluxes}. 
Each panel shows the normalized H$\alpha$ flux versus normalized total flux for one of the four observed flares. Data are separated into three temporal phases: pre-flare (green), flare (orange), and post-flare (blue). Linear fits are shown for the flare phase (orange line) and for the combined pre- and post-flare data (blue line), with shaded regions indicating $\pm 1\sigma$ confidence intervals. Pearson correlation coefficients ($R$) are listed in matching colors. There is a tight correlation between H-alpha flux and total flux during the flare, and this correlation breaks down in the pre- and post-flare periods. This decoupling supports our conclusion that  the post-flare flux enhancement is not due to a lingering flare decay. 
}
\label{fig:h_alpha_tot_flux_correlations}
\end{figure*}

Here, we demonstrate that the spectra of the flare and the post-flare enhancement are clearly distinct.

We begin with the observed JWST/NIRISS spectra, averaging them over three key time windows: pre-flare, maximum flare light, and post-flare (see color-coded markers in Fig.~\ref{fig:datasets_and_offstes}). Figure~\ref{fig:spectral_variations} shows the differences between the maximum-light and post-flare spectra relative to the pre-flare baseline, highlighting the spectral changes at maximum flare light and in the post-flare phase. At maximum flare light, the spectrum shows a strong flux increase across the entire NIRISS range (0.6–2.8~$\mu$m), with the largest increases at short wavelengths. These features are consistent across all four events reflecting a strong flare heating of the atmosphere. In addition to the hot continuum, prominent emission lines appear during the flare, including H$\alpha$, He~I ($\lambda$10830), and hydrogen Paschen and Brackett series lines—well-established indicators of strong flare activity in both solar and stellar contexts \citep{Benz2017, Kowalski2024, Howard2023}. These emission lines are shown in detail in Figure~\ref{fig:emission_lines_normalized} and discussed further in Appendix~\ref{app:emission_lines}.

In contrast, the post-flare spectra exhibit no detectable emission lines (see Fig.~\ref{fig:spectral_variations} and Appendix Fig.~\ref{fig:emission_lines_normalized}), and the continuum appears flat or slightly elevated in the 0.85–1.8~$\mu$m range. This suggests that the hot component responsible for the flare emission has dissipated, and the temperature structure of the atmosphere has returned to a cooler, quiescent state. A weak H$\alpha$ emission is detected only in Flare~2 (see also Fig.~6 in Appendix~\ref{app:emission_lines}), likely due to the short ($\sim$1.5 hr) interval between the flare peak and the post-flare window or possibly due to a secondary flare superimposed on the decay phase.

One possible explanation for the post-flare flux enhancement is a lingering flare. However, while late-phase flare continua cool significantly over time \citep{Kowalski2013, Kowalski2016}, they are typically accompanied by persistent line emission, especially in H$\alpha$. For example, \cite{Kowalski2013} (see their Figs.~14 and 15) show that H$\alpha$ emission remains elevated even after the flare continuum has faded. In contrast, our JWST/NIRISS observations reveal no detectable emission lines—including H$\alpha$—more than 2.5 hours after the flare peak. The measured light curves also flatten, indicating that the observed enhancement is not accompanied by the gradual decay expected from a lingering flare. Taken together, these factors make a lingering flare an unlikely explanation for the post-flare flux enhancement.
Instead, it likely marks the disappearance of a (part of a) magnetic feature associated with the flares. Consequently, the blue curves in Figure~\ref{fig:spectral_variations} represent the difference in the spectral energy distribution of this feature relative to the quiet star. 
Indeed, after the flare, a part of the feature is replaced by a quiet photosphere. Therefore, the difference between post- and pre-flare spectra reflects the contrast between the quiet photosphere and the magnetic feature. The fact that the difference spectrum is positive across most wavelengths implies that the magnetic feature was darker than the quiet photosphere.

\subsection{Relationship between H$\alpha$ and total flux}\label{sec:correcltions}
We show that the relationship between H$\alpha$ and total flux differs significantly between the flare and non-flare phases. We quantify this relationship by comparing their correlation across three intervals: pre-flare, flare peak, and post-flare (Fig.~\ref{fig:h_alpha_tot_flux_correlations}).

During the flare phase, linear fits to the H$\alpha$–total flux relation show steep slopes, reflecting a strong and coherent response in both emission lines and continuum. This is quantitatively supported by the Pearson correlation coefficients, which are high across all four events ($R \approx 0.89$–$0.98$). In contrast, the pre- and post-flare phases exhibit much shallower or nearly flat slopes. Specifically, post-flare measurements in Flares 1, 3, and 4 show weak or slightly negative correlations ($R \sim -0.2$ to $-0.1$), indicating that the flux enhancement is not a continuation of the flare. Only Flare 2 shows a modest post-flare correlation ($R \approx 0.2$), consistent with its short post-flare baseline and a small subsequent flare after the maximum light of the main flare. These results, combined with the absence of line emission and hot continuum, support the interpretation that the flux enhancement reflects a surface transformation, likely the disappearance of an  active region or a part thereof.

\subsection{Areas and spectra of vanished magnetic features}\label{sec:areas}
From the observed relative change in total flux, we estimated the projected areas of the active features that disappeared. These estimates depend on the assumed contrast between the active region and the quiet stellar surface. Because the change in observed flux reflects both contrast and area, there is an inherent degeneracy -- darker regions produce the same signal with less surface coverage \citep{Pettersen1992}. To explore plausible scales, we considered three illustrative cases: a completely dark region (black spot), an umbra, and a penumbra.

\begin{figure*}
\centering
\includegraphics[width=1.0\textwidth]{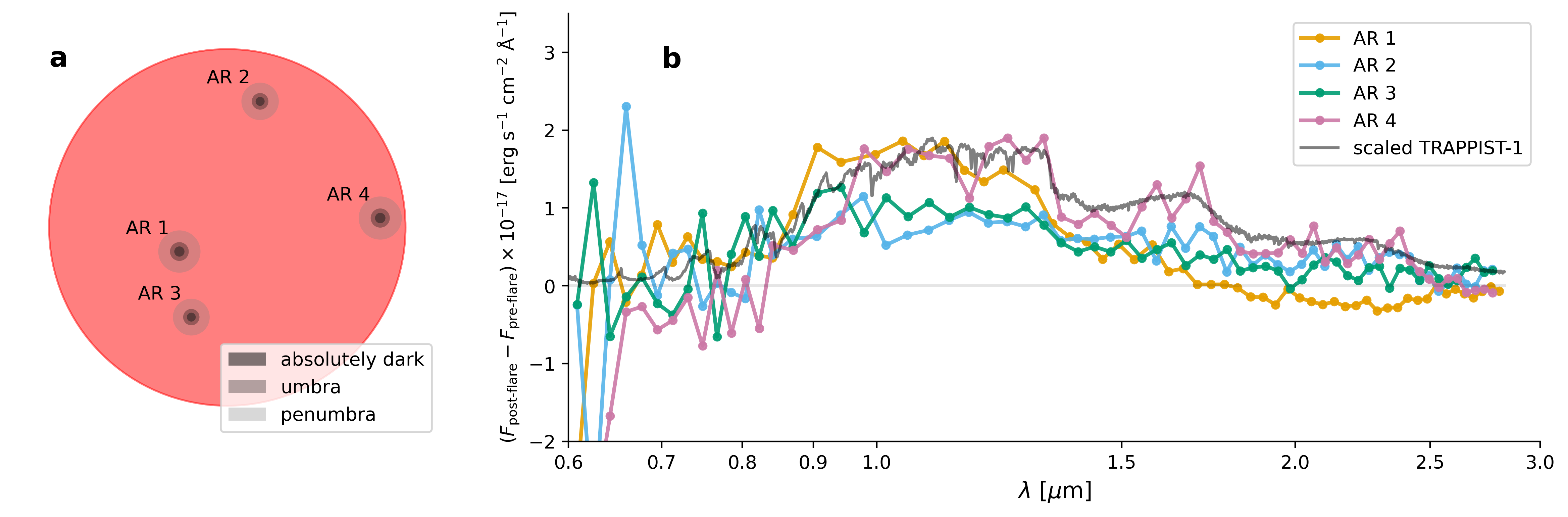}
\caption{\textbf{Magnetic surface features vanishing during flares on TRAPPIST-1.}
a)  Schematic representation of the inferred projected surface areas of magnetic features that disappeared during four flare events on TRAPPIST-1. The colored circles illustrate the projected surface areas required to explain the observed post-flare flux enhancement, assuming three limiting cases for the brightness contrast of the vanished region: a completely black spot (darkest), an umbra-like region (medium gray), and a penumbra-like region (light gray). Limb darkening, spatial distribution, and foreshortening are not considered. 
 We note that these projected surface areas represent only the vanished part of the active region. The total size of the original active region may be larger.
b) Spectra of the vanishing active regions shown in panel (a).
Each colored curve shows the difference between the post-flare and pre-flare spectra for one of the four analyzed flares (AR 1–4), interpreted as the spectral energy distribution of a magnetic active surface feature that vanished during the flare. The spectra are binned over 40 wavelength intervals for clarity.
The black curve shows the quiescent spectrum of TRAPPIST-1, obtained by time-averaging the data during the TRAPPIST-1f visit. The spectrum is scaled for comparison.  
The figure shows that the spectral profile of the vanished magnetic features does not differ substantially from that of TRAPPIST-1.
}
\label{fig:active_region_sizes_and_specs}
\end{figure*}

For the umbra and penumbra scenarios, we used spectral contrasts derived from synthetic spectra based on a MURaM \citep{Volgler2005, Witzke2024} 3D MHD simulation of an M4-type star spot (Bhatia et al. in prep) -- the coolest available model -- processed with the radiative transfer \texttt{MPS-ATLAS} code \citep{Witzke2021}. While TRAPPIST-1 is significantly cooler (M8), these contrasts provide a reasonable lower bound for active region brightness, as both umbral and penumbral contrasts tend to decrease with decreasing stellar temperature \citep{Berdyugina2005, Smitha2025}. Under these assumptions, the inferred projected areas are $0.06-0.09$\% (completely black),  $0.19-0.29$\% (umbra-like), and  $1.0-1.5$\% (penumbra-like) of the visible stellar disk. These three scenarios and the corresponding projected areas are visualized in Figure~\ref{fig:active_region_sizes_and_specs}a. 
The estimated surface coverage of the vanished features is consistent with the estimates of spot sizes reported by \citet{Rathcke2025}, which were inferred from simultaneous modeling of contamination of TRAPPIST-1b transmission spectra and TRAPPIST-1 spectral variability.

The spectra of the vanishing magnetic features (AR 1–4)\footnote{Original unbinned spectra of the vanishing active features are available online at Zenodo \cite{vasilyev2025zenodo}.}  reveal a clear energy maximum near the peak of the spectral energy distribution of TRAPPIST-1 itself, suggesting that these disappearing magnetic regions cannot be significantly cooler than the stellar photosphere (see Fig.~\ref{fig:active_region_sizes_and_specs}b). 
In Appendix C, we estimate the temperatures of the disappearing magnetic features to be between 2350 K and 2520 K. These estimates are derived by fitting a Planck function to the spectra of the magnetic features. As expected, the spectra are not well described by a Planck function, so the estimates should be considered illustrative. 

\section{Conclusions}\label{sec:concl}
We report the detection of post-flare flux enhancements in JWST/NIRISS spectroscopy of the cool M8 dwarf TRAPPIST-1.  These flux offsets are spectrally distinct from the flare and lack emission lines typically associated with stellar flares. We interpret them as the spectroscopic signature of a (partially) disappearing magnetic feature -- analogous to a
phenomenon that has been directly observed on the Sun.  
By differencing pre- and post-flare spectra, we extract the spectrum of the vanished feature, providing, to our knowledge, the first direct measurement of the spectral energy distribution of  the magnetic features on TRAPPIST-1. Remarkably, the extracted spectra are relatively similar to the stellar spectrum itself, indicating that the disappearing magnetic features  were only slightly cooler than the quiet stellar surface, with effective temperatures of approximately 2350–2550~K. 

Previous ground-based spectroscopic flare studies lacked the wavelength coverage, photometric stability, and sensitivity required to detect faint, broadband infrared excesses. In this context, JWST provides access to an observational regime that was previously unexplored, enabling the detection of flare-induced stellar surface transformations.

We cannot fully exclude that this flux enhancement is the long-lived flare decay, but the absence of emission lines, the stable post-flare plateau, and the breakdown of the total flux correlation with the H$\alpha$ flux all argue for a separate surface phenomenon rather than a tail of the flare. While the observational signatures point to the disappearance of a dark magnetic feature, its exact nature remains uncertain. In particular, the available data do not allow us to determine whether the entire magnetic feature disappears or only a part of it, as is often observed in the solar case.

Our approach opens a new pathway for characterizing magnetic features on stars beyond the Sun using space-based spectroscopy.

\begin{acknowledgments}
VV and NK acknowledge support from the Max Planck Society under the grant ``PLATO Science'' and from the German Aerospace Center under ``PLATO Data Center'' grant   50OO1501. The study was supported by the
European Research Council (ERC) under the European Union's
Horizon 2020 research and innovation program (grant No.
101118581---project REVEAL). 
This material is based upon work supported by the National Aeronautics and Space Administration under Agreement No.\ 80NSSC21K0593 for the program ``Alien Earths''.
The results reported herein benefited from collaborations and/or information exchange within NASA’s Nexus for Exoplanet System Science (NExSS) research coordination network sponsored by NASA’s Science Mission Directorate.
Based on observations with the NASA/ESA/CSA James Webb Space Telescope obtained from the Mikulski Archive for Space Telescopes (MAST) at the Space Telescope Science Institute, which is operated by the Association of Universities for Research in Astronomy, Incorporated, under NASA contract NAS5-03127. Support for program number JWST-AR-05370 was provided through a grant from the STScI under NASA contract NAS5-03127.
\end{acknowledgments}

\software{
matplotlib \citep{matplotlib}, 
numpy \citep{numpy}, 
scipy \citep{scipy}
}

\appendix
\section{NIRISS/SOSS Datasets} ~\label{app:data}

\begin{deluxetable*}{cccccccc}
\tablecaption{Summary of NIRISS/SOSS Transit Observations of TRAPPIST-1 Planets\label{tab:obs_summary}.}
\tablehead{
\colhead{Planet} &
\colhead{Program} &
\colhead{PI} &
\colhead{Obs.\ ID} &
\colhead{Start Date (UT)} &
\colhead{Duration (hr)} &
\colhead{Flare num.} &
\colhead{Reference}
}
\startdata
b & GO\,2589  & Lim        & 1   & 2022-07-18 & 6.47 & & \citet{Lim2023} \\
b & GO\,2589  & Lim        & 2   & 2022-07-20 & 6.47 & & \citet{Lim2023} \\
c & GO\,2589  & Lim        & 3   & 2022-10-28 & 6.67 & & \citet{Radica2025} \\
c & GO\,2589  & Lim        & 4   & 2023-10-31 & 6.76 &  1 & \citet{Radica2025} \\
f & GTO\,1201 & Lafreni\`ere & 101 & 2022-10-28 & 5.03 &  2 & Lim et al.\ (in prep) \\
f & GTO\,1201 & Lafreni\`ere & 105 & 2023-06-15 & 5.13 & & Lim et al.\ (in prep) \\
f & GTO\,1201 & Lafreni\`ere & 104 & 2023-06-24 & 5.13 & & Lim et al.\ (in prep) \\
f & GTO\,1201 & Lafreni\`ere & 102 & 2023-07-03 & 5.13 &  3 & Lim et al.\ (in prep) \\
f & GTO\,1201 & Lafreni\`ere & 103 & 2023-07-22 & 5.13 &   4 & Lim et al.\ (in prep) \\
\enddata
\tablecomments{
All observations were obtained in NIRISS SOSS mode using the SUBSTRIP256 subarray and the NISRAPID readout pattern with 18 groups per integration.
Observational details are summarized from TrExoLiSTS \citep{Nikolov2022}.
}
\end{deluxetable*}

This study utilizes the complete set of JWST NIRISS/SOSS observations (0.6--2.85\,\micron{}, $R{\sim}700$) of the TRAPPIST-1 system available in the MAST archive as of early 2025. 
These observations span two programs---GO\,2589 (PI: Lim) and GTO\,1201 (PI: Lafrenière)---and include a total of nine transits covering planets b, c, and f. 
A summary of the visits is provided in Table~\ref{tab:obs_summary}.

From GO\,2589, we include two transits of TRAPPIST-1b (2022 July 18 and 20) and two transits of TRAPPIST-1c (2022 October 28 and 2023 October 31). 
The TRAPPIST-1b observations were analyzed in \citet{Lim2023}, who identified strong stellar contamination from unocculted spots and faculae. 
The TRAPPIST-1c visits were presented in \citet{Radica2025}, who jointly modeled stellar and planetary components to constrain the planet's atmospheric composition.

In addition, we include five transits of TRAPPIST-1f observed as part of GTO Program 1201 (on 2022 October 28, and 2023 June 15, June 24, July 3, and July 22), each with similar observational configurations and durations of approximately 5.1 hours. 
These data are currently the subject of ongoing analysis (Lim et al., in prep). 
For all datasets, we analyze the reduced time-series spectra provided by Olivia Lim (priv.\ comm.), which were processed using the data reduction procedures described in \citet{Lim2023}.

\section{Flare-Induced Emission Lines} \label{app:emission_lines}
\begin{figure*}[ht!]
    \centering
\includegraphics[width=1.0\textwidth]{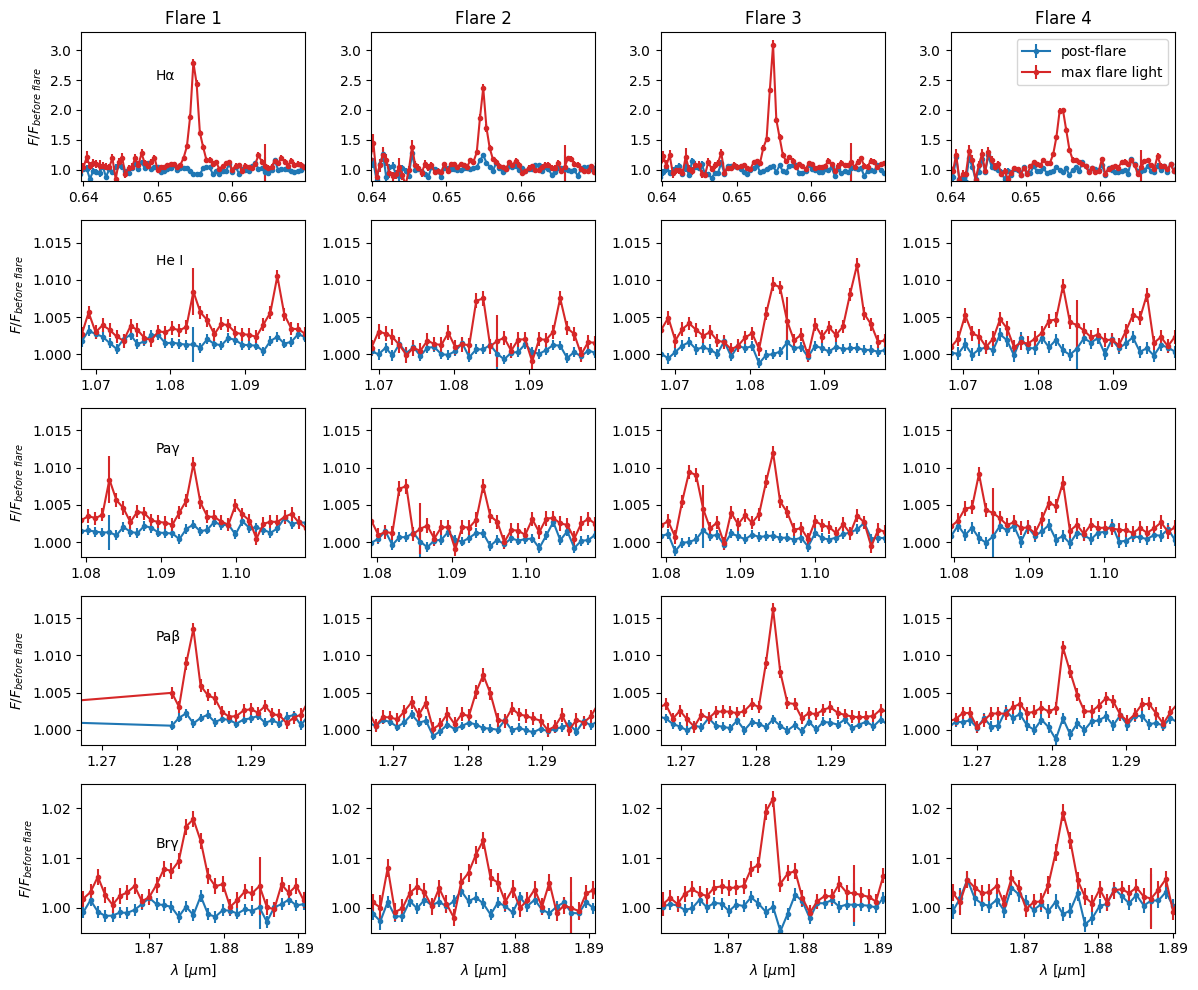}
    \caption{Line profiles of flare-induced emission features in normalized JWST/NIRISS spectra.
    Each column corresponds to a different flare event. Rows show zoomed-in views of selected spectral lines typically associated with flare activity: the hydrogen Balmer (H$\alpha$), Paschen (Pa$\gamma$, Pa$\beta$), and Brackett (Br$\gamma$) series, as well as the He I triplet at 1.083 $\mu$m. The red curves represent spectra averaged at the flare peak, and the blue curves show post-flare spectra, both normalized to the pre-flare level.}
    \label{fig:emission_lines_normalized}
\end{figure*}
For each dataset, we selected a time window of duration $\Delta t \approx0.08$~h during the maximum flare brightness and $\Delta t \approx0.26$~h during the pre-flare and post-flare phases. Averaged spectra were computed for each interval, and then the flare and post-flare spectra were normalized by the pre-flare spectrum (see Figure~\ref{fig:emission_lines_normalized}). The  flare spectra exhibit strong emission lines of Ca~II (0.85~$\mu$m), He~I (1.083~$\mu$m) and several hydrogen lines: Pa$\gamma$ (1.094~$\mu$m), Pa$\beta$ (1.282~$\mu$m), and Br$\gamma$ (2.166~$\mu$m) at the time of the strongest H$\alpha$ emission. These lines are most pronounced during Flare~1 and~3.
 In the normalized post-flare spectra, a weak H$\alpha$ emission is detected only in Flare~2, consistent with the short ($\sim$1.5 hr) delay between the flare peak and the post-flare window. 
 This suggests the post-flare spectrum was taken during the decay phase of  the flare. In the other three datasets, no emission lines are visible, indicating that the stellar atmosphere had returned to a quiescent state.

\section{Temperatures of vanished  active features.} 
\label{app:temperatures_of_active_regions}
\begin{figure*}[ht!]
    \centering
    \includegraphics[width=1.0\textwidth]{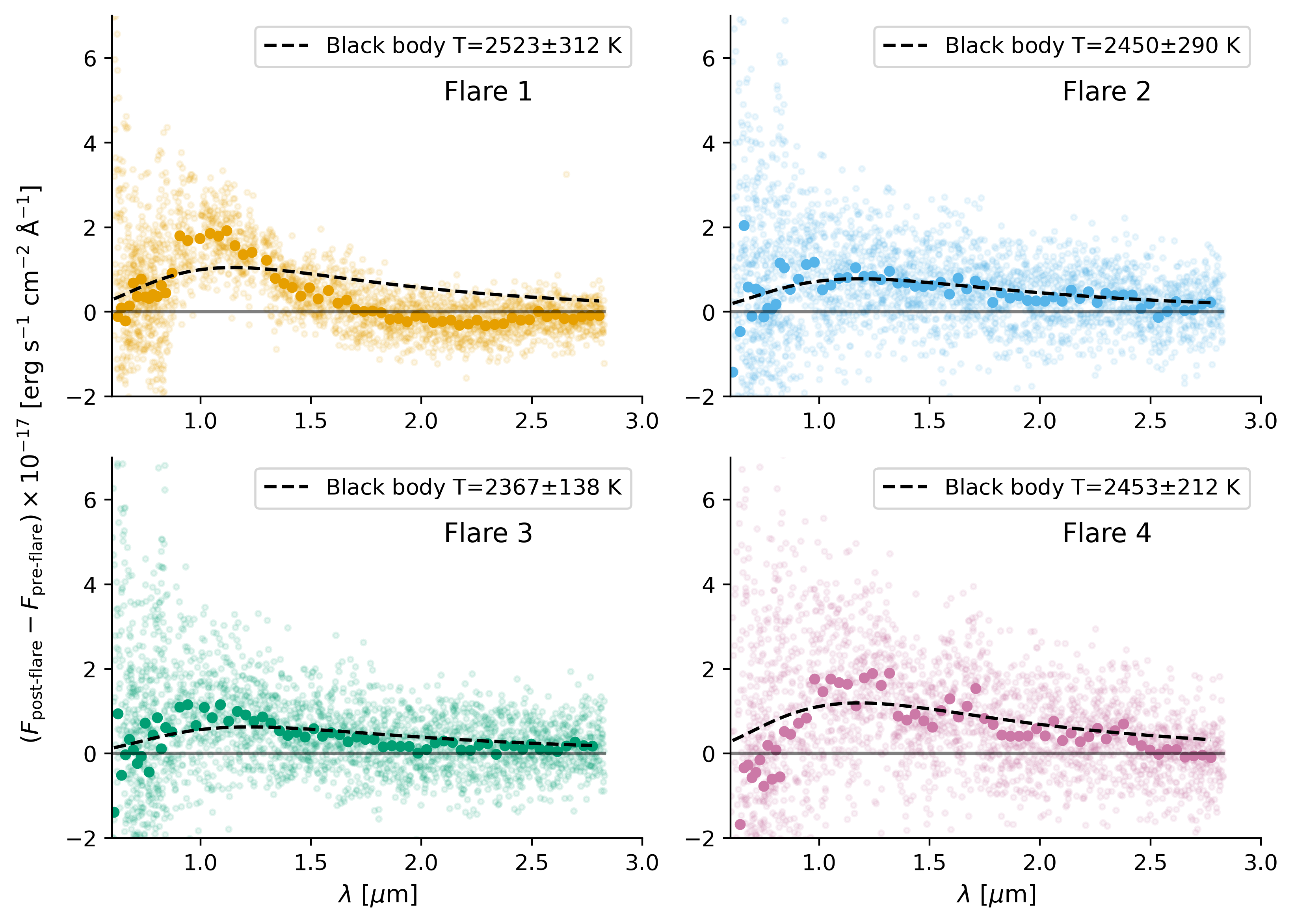}
    \caption{Temperature estimates of the vanishing dark features.
    Each panel shows the flux difference between post- and pre-flare spectra, interpreted as the spectral energy distribution of the vanished active region (binned over 40 points; bold markers). 
    The black dashed line shows the Planck function that best fits this spectrum, used to estimate the temperature of the vanished region. The temperature and its 1$\sigma$ uncertainty are indicated in the legend. }
\label{fig:specs_acive_regions_bb_fit}
\end{figure*}
We assume that the spectral energy distribution of the active region is described by a Planck function. The spectrum of the vanished active region is obtained as the difference between the pre-flare and post-flare spectra. For each event, we fit this model using temperature and normalization as free parameters. The fits are performed on data binned to the NIRISS resolution (each bin represents the average of 40 original spectral points), using 200 bootstrap trials. In each trial, 50–70\% of the binned points are randomly selected, following the sampling strategy of \citet{Howard2023}, to reduce sensitivity to local spectral fluctuations. In Figure~\ref{fig:specs_acive_regions_bb_fit}, we show the original and binned data alongside the best-fitting Planck curves. The inferred temperatures of the active regions range from 2367 K to 2523 K, with a typical uncertainty of a few hundred kelvins. These values are not significantly cooler than the stellar surface temperature of TRAPPIST-1 \citep{VanGrootel2018, Gonzales2019, Davoudi2024}, as expected from the relatively small temperature contrasts between active regions and the photosphere in M-type stars \citep{Berdyugina2005, Smitha2025}.

\bibliography{main}{}
\bibliographystyle{aasjournalv7}



\end{document}